\documentclass[12pt]{iopart}
\usepackage{color}
\usepackage{float}
\usepackage{graphicx}
\usepackage{bm}

\newcommand{\half}{\mbox{$\frac{1}{2}$}}
\begin{document}

\title{L$_1$ Regularization for Reconstruction of a non-equilibrium Ising Model}

\author{Hong-Li Zeng$^1$, John Hertz$^{2,~3}$ and Yasser Roudi$^{2,~4}$}

\address{$^1$ Department of  Applied Physics, Aalto University, FIN-00076 Aalto, Finland\\
$^2$ Nordita, KTH Royal Institute of Technology and Stockholm University, Roslagstullsbacken 23, SE-106 91, Stockholm, Sweden\\
$^3$ The Niels Bohr Institute, 2100 Copenhagen, Denmark\\
$^4$ Kavli Institute for Systems Neuroscience, NTNU, 7030 Trondheim, Norway }
\eads{\mailto{hongli.zeng@aalto.fi}, \mailto{hertz@nbi.dk} and \mailto{yasserroudi@gmail.com}}

\begin{abstract}
The couplings in a sparse asymmetric, asynchronous Ising network are reconstructed using an exact learning algorithm.   L$_1$ regularization is used to remove the spurious weak connections that would otherwise be found by simply minimizing the minus likelihood of a finite data set.   In order to see how L$_1$ regularization
works in detail,  we perform the calculation in several ways including (1) by iterative minimization of a cost function equal to minus the log likelihood of the data plus an L$_1$  penalty term, and (2) an approximate scheme based on a quadratic expansion of the cost function around its minimum.  In these schemes, we track how connections are pruned as the strength of the L$_1$ penalty is
increased from zero to large values.  The performance of the methods
for various coupling strengths is quantified using ROC curves.

\end{abstract}

\maketitle

\section{Introduction}

%----\textbf{Background:}
A crucial step in understanding how a complex network operates is inferring
its connectivity from observables in a systematic and controlled way. This learning
of the connections from data is an inverse problem. Recently, with the
ongoing growth of available data, especially in biological systems,
such inverse problems have attracted a lot of attention in
statistical physics community. Examples of applications
include the reconstruction of a gene regulation network from gene expression
levels \cite{bailly2010inference} and identification of the protein-protein
interactions from the correlations between amino acids \cite{weigt2009identification}.
One proxy for such a problem is the inverse Ising model, where the parameters
of the model (fields and interactions) are inferred from observed spin history.

%----\textbf{Here,dynamical problem:$J_{ij}\neq J_{ji}$}
There has been a long history of inferring Gibbs equilibrium models, such as the
equilibrium Ising model, where the fields and couplings are inferred from the measured
means and correlations \cite{Ackley85,Kappen98,schneidman2006weak,roudi20092}.
The methods developed for learning the connections in these models as originally
formulated, do not assume any prior belief about the network architecture and
they only use the data to decide about that. Recently, though it has been
shown that the connections can be inferred much more efficiently when a sparse
prior, specifically the L$_1$ regularizer, is taken into account \cite{ravikumar2010high,aurell2011inverse}.
However, our theoretical understanding of how L$_1$ regularization works is limited.

In many practical applications, equilibrium models would be of limited use.
For instance, most biological systems operate in out-of-equilibrium regimes.
Consequently, equilibrium models are not usually good candidates to infer
the interactions and may fall short even as generative models for describing the
statistics of the data \cite{roudi20091}.
Several recent studies have thus moved to kinetic models, prescribing
exact and approximate learnings for inferring the
connections in non-equilibrium models \cite{roudi2011mean, hertz2010inferring, zeng2011network}.
However this body
of work has not yet exploited the potential power of L$_1$ regularization in inferring the connections.

In this paper, focusing on the asynchronously updated Ising model, we will describe how
L$_1$ regularization helps in inferring the connections in a non-equilibrium model.
We try to shed light on the mechanics by which the regularization works through
developing approximate ways of performing L$_1$. We study how the regularization
shrink the connections gradually with the increase of regularization parameter.

%---\textbf{Something about using a sparse model}

%----\textbf{Organization of the paper}
The paper is organized as follows: The dynamics and the underlying network are described in section \ref{sec2}, an L$_1$-regularized learning rule for an asynchronously updated kinetic Ising model is described in section \ref{sec3}, approximate learning algorithms, based on an expansion of the cost function of section \ref{sec3}, are derived in section \ref{sec4}, and the performance of the learning rules is studied in section \ref{sec5}. The effects of different coupling strengths are explored in section \ref{sec6}. A discussion is given in section \ref{sec7}.

\section{Glauber dynamics and network}\label{sec2}
%We collect the observable $m_i=\langle s_i(t)\rangle_t$ and $C_ij(\tau)=\langle(s_i(t+\tau)-m_i)(s_j(t)-m_i)\rangle$ from kinetic Ising model, which was originally motivated by neural networks.

We consider a kinetic Ising model endowed with Glauber dynamics  \cite{glauber1963time}.
Glauber dynamics describes the evolution of the joint probability of the spin states
$p(S_1, S_2, ... , S_N;t)$ in time $t$, following the master equation

\begin{equation}\label{1}
\frac{dp(\textbf{s};t)}{dt}=\sum_i\omega_i(-s_i)p(s_1,...,-s_i,...,s_N;t)
-\sum_i\omega_i(s_i)p(\textbf{s};t).\\
\end{equation}
where,
$$\omega_i(s_i(t))=\frac{\gamma_0}{1+\exp\left[2s_i(t)H_i(t)\right]}=\frac{\gamma_0}{2}\left[1-s_i(t)\tanh H_i(t)\right],$$
is the probability for spin $i$ to change its state
from $S_i(t)$ to $-S_i(t)$ during the time interval $dt$. Here, we choose time units so
that $\gamma_0=1$. The quantity $H_i(t) = h_i +\sum_jJ_{ij}s_j(t)$ is the instantaneous
field acting on spin $i$.  The external field $h_i$ can be dependent on time, but for the
sake of simplicity we focus on the stationary case, i.e., time-independent $h_i$, here.

One way to implement the Glauber dynamics is as follows: Make a discretization of the evolution process with very small time steps $\delta t\ll 1/N$.   At
each step, every spin is selected for updating with probability $\delta t$. For $\delta
t\ll 1/N$, almost certainly only one spin at a time will be updated. The next value of the
spin selected for updating is chosen according to
\begin{equation}\label{eq:update}
p(s_i(t+\delta t)|s_i(t))=\frac{\exp [s_i(t+\delta t)H_i(t)]}{2\cosh H_i(t)}
= \half [ 1 + s_i(t+\delta t) \tanh H_i(t)].
\end{equation}
Note that the updated spin might not change its value; an update is not necessarily a flip.
In this paper, we will take the dynamics to be defined in this doubly stochastic way and assume that the data accessible to us include both the times at which every spin is selected for updating (determined by an independent Poisson processes for each spin) and the result of those updates (whose outcomes are given by (\ref{eq:update}), i.e., the spin history).  The problem may also be treated by other algorithms that only assume knowledge of the spin history (not of all the update times); these are discussed in other work \cite{Zeng2012algorithms}, but we do not consider them here. In our computations, in order not to waste lots of time not updating any spins, we have, at each time step, chosen exactly one spin at random for updating. For finite $N$ this is not exactly the dynamics described above, but we do not see any difference when we compare the results of our computations with those done following the correct dynamics exactly.

We study a diluted binary asymmetric Sherrington-Kirkpatrick (SK) model with these
dynamics. For the original SK model, the pairwise interactions $J_{ij}$ between
spins $i$ and $j$ were i.i.d. Gaussian variables (except $J_{ij} = J_{ji}$) with variance $g^2/N$ and mean 0.   In the model we study here, the network is diluted, $J_{ij}$ is independent of $J_{ji}$, and the interactions vary only in sign, not in magnitude: Each coupling has the distribution
\begin{equation}\label{eq:structure}
p(J) = \frac{c}{2N}\delta \left (J-\frac{g}{\sqrt{c}}\right)
+ \frac{c}{2N}\delta \left (J+\frac{g}{\sqrt{c}}\right)
+ \left( 1-\frac{c}{N}\right)\delta(J).
\end{equation}
where $c$ is the average in-degree (and out-degree). We are interested in sparse networks, i.e., $c\ll N$.  In our computations, we use $N=40$ and $c=5$. Furthermore, as mentioned above, we model asymmetrically coupled spins, taking each $J_{ij}$ independent of $J_{ji}$.  This model can have a stationary distribution (and does for the parameters we use here), but it is not of Gibbs-Boltzmann form, and no simple expression for it is known.

\section{Exact learning}\label{sec3}

As described above, we suppose we know the full history of the system -- both the
$\{s_i(t)\}$, with $1\leq i\leq N$ and $1\leq t\leq L$, where $L$ is the data length, and
the update times $\{\tau_i\}$.  We can reconstruct the couplings $J_{ij}$ and external
fields $h_i$ by performing the gradient descent on the negative log-likelihood of this history, which is given by
\begin{equation}
-{\cal L}_0   = -\sum_i \sum_{\tau_i}\left[ s_i(\tau_i+\delta t)H_i(\tau_i) - \log 2 \cosh H_i(\tau_i)\right].
\label{eq:L0}
\end{equation}
We can minimize the log-likelihood by simple gradient descent with a learning rate $\eta$:
\begin{equation}
\delta J_{ij} = \eta \frac{\partial
{\cal L}_0}{\partial J_{ij}}= \eta
\sum_{\tau_i} [s_i(\tau_i+\delta t)-\tanh H_i(\tau_i)]s_j(\tau_i).	\label{eq:learning_noL1}
\end{equation}
This equation includes the learning rule for the external field $h_i$ under the convention
$J_{i0} = h_i$, $s_0(t) = 1$.  It has the same form as that for a synchronous model,
except that changes for spin $i$ are made only at times $\tau_i$.

For finite $L$, this procedure will in general produce a densely-connected network.  To sparsify it, we add a simple regularization term that penalizes dense connectivity in a
controllable fashion. We then minimize a cost function
\begin{equation}\label{eq:cost}
E=-{\cal L}_0+\Lambda\sum_{ij}|J_{ij}|.
\end{equation}
where the first term is the negative log-likelihood and the second term is the $L_1$ norm. There are several efficient methods have been used to minimize the cost function (\ref{eq:cost}), e.g.,
the interior-point method \cite{Interior-point1, interior-point2}. However,
in order to see how L$_1$ regularization works in detail, we study a simple
gradient descent algorithm here.   Gradient descent on this cost function leads
to an additional term in the learning rule for couplings:
\begin{equation}
\label{eq:learning_rule}
\delta J_{ij}=\eta_J\left\{\sum_{\tau_i}\left[s_i(\tau_i+\delta t)-\tanh H_i(\tau_i)\right]s_j(\tau_i) -\Lambda {\rm sgn}(J_{ij})\right\}.
\end{equation}

The log-likelihood function $\mathcal{L}_0$ is smooth and convex as a function of the $J_{ij}$ and $h_i$, so the cost function is concave except on the hyperplanes where any $J_{ij}=0$.  This leads to complications in the minimization whenever a minimum of $E$ is at $J_{ij}=0$:  We deal with this problem by setting $J_{ij}=0$ whenever the change (\ref{eq:learning_rule}) would cause $J_{ij}$ to change sign. Then, if the minimum of $E$ truly lies at this $J_{ij}=0$, the estimated $J_{ij}$ will oscillate between zero and a small nonzero value (using ${\rm sgn}(0) = 0$).   However, the size of these oscillations is proportional to the learning rate $\eta$, so a sufficiently small $\eta$ ensures that these couplings can be pruned by a simple rounding procedure, with negligible chance of removing coupling that are not truly zero at the minimum.  In the case that $J_{ij}$ is not zero at the minimum, its estimated value will continue to change and it will move toward its optimal value after the step where it was set to zero.

Another way to deal with the non-differentiability of the cost function (\ref{eq:cost}) is to use
$\Lambda\mu\sum_{i,j}\log\cosh(J_{ij}/\mu)$ as the penalty term and take the limit
$\mu\rightarrow0$. This term leads to the replacement of the $\lambda {\rm sgn}(J_{ij})$ by $\lambda\tanh(J_{ij}/\mu)$. For any non-zero $\mu$, this modified cost function is totally convex. We checked some of our computations by doing the regularization this way. No difference between these results and those done as described above was found.

\section{An approximate learning scheme}\label{sec4}

We can get some insight into the dynamics of the learning with regularization by expanding
the cost function (\ref{eq:cost}) to second order around its minimum  $\mathbf{J^0}$ when
$\Lambda=0$.  Up to a constant, we have
\begin{equation}\label{eq:cost1}
\frac{E}{T}= \half \sum_{ijk}C^{(i)}_{jk}v_{ij}v_{ik}+\lambda\sum_{ij}|J_{ij}^0+v_{ij}|
\end{equation}
where $v_{ij}=J_{ij}-J^0_{ij}$, $T=L/N$ is the number of updates per spin, $\lambda=\Lambda/T$, and
\begin{equation}
C^{(i)}_{jk}=\frac{1}{T}\sum_{\tau_i}(1-\tanh^2H_i^0(\tau_i))\delta s_j(\tau_i)\delta s_k(\tau_i).
\label{eq:Fisher1}
\end{equation}
Since the quantities in the sum in (\ref{eq:Fisher1}) are insensitive to whether spin $i$ is updated, the average over updates may safely be replace by an average over all times,
\begin{equation}
C^{(i)}_{jk}= \langle (1-\tanh^2H_i^0(t))\delta s_j(t)\delta s_k(t) \rangle_t
\label{eq:Fisher2}
\end{equation}
the Fisher information matrix for spin $i$, which is a more robust quantity.

Minimizing (\ref{eq:cost1}), we get, to first order in $\lambda$,
\begin{equation}
\sum_{k}C^{(i)}_{jk}v_{ik}=-\lambda{\rm sgn}(J_{ij}^0+v_{ij})\approx-\lambda {\rm sgn}(J_{ij}^0).
\end{equation}
Solving this equation for $v_{ij}$,we obtain:
\begin{equation}\label{Fisher}
v_{ij}=-\lambda\sum_{k}\left[C^{(i)}\right]^{-1}_{jk}{\rm sgn}(J_{ik}^0).
\end{equation}
This equation shows how the regularization term shrinks the magnitudes of the couplings.

In the weak coupling limit (small $g$ or, equivalently, high temperature), $\left[C^{(i)}\right]^{-1}_{jk} = \delta_{jk}$, so the $J_{ij}$ are just shrunk in magnitude proportional to $\lambda$ until they reach zero and are pruned.  This is a trivial kind of regularization: We know that the couplings that survive the pruning procedure the longest are simply the ones with the biggest initial absolute values.  In this case, there is no need to go through the elaborate learning-with-regularization procedure of (\ref{eq:learning_rule}).  However, at larger coupling this is not the case.  Some $J_{ij}$ will be shrunk more rapidly than others, depending on the size and signs of the terms in the sum in (\ref{Fisher}).

Based on the quadratic expansion (\ref{eq:cost1}), we can carry out the pruning in an approximate alternative fashion, as follows:  Starting from $J_{ij}^0$ and a small value of $\lambda$, we calculate the shifts $v_{ij}$ by (\ref{Fisher}) and remove any $J_{ij}$ that would go though zero.  Starting from the resulting new $J_{ij}$s (some of them now equal to zero), increase $\lambda$, recalculate the Fisher information matrix and calculate new shifts in the parameter values.  Again remove any couplings that change sign, and continue until the desired degree of pruning has been achieved.  This amounts to numerical integration of the differential equation, describing a kind of dynamics of regularization under increasing $\lambda$.
\begin{equation}
\frac{dJ_{ij}(\lambda)}{d\lambda} = - \sum_{k}\left[C^{(i)}(\lambda)\right]^{-1}_{jk}{\rm sgn}(J_{ik}(\lambda)).			
\label{eq:dJdlambda}
\end{equation}
Note that this procedure requires only equal-time average quantities, unlike the full computation following the learning rule (\ref{eq:learning_rule}).

If one knows {\em a priori} what value of $\lambda$ to use, it is probably not an advantage to use this algorithm.  One can simply do the full computation once, at that value, while with this approximate algorithm we have to simulating the model to estimate the Fisher matrices at all the intermediate $\lambda$s in integrating (\ref{eq:dJdlambda}).  On the other hand, one may not know the optimal $\lambda$.  It then becomes necessary to explore the regularized model over some wide range of $\lambda$.   In this case, the approximate algorithm will have a speed advantage, because it only requires a learning loop at the initial $\lambda$ (zero, in the case described here).

\section{Results}\label{sec5}
We consider the problem of identifying the positive and negative couplings in the network,
i.e., correctly classifying every potential bond as $+$, $-$ or 0. Consider first the
couplings $J_{ij}$s found with no regularization, i.e., $\lambda=0$. For given $g$, $c$
and $N$, for very large $T$ the inferred $J_{ij}$ will be very close to the true ones.  A
histogram of their values will have three narrow peaks around $0$ and $\pm g/\sqrt{c}$,
and it will be trivial to identify the true nonzero couplings and their signs (figure~1a,b).
 In the opposite limit (small $T$), the data are not sufficient to estimate the couplings
well.  The histogram will be unimodal, and it will be more or less hopeless to solve the
problem, even with the help of L$_1$ regularization (figure~1c,d).  The interesting case is
that of intermediate data length, for which the partial histograms from the zero and
nonzero-J classes overlap, but the separations between their means are not much smaller
than their widths (figure~1e,f). We would also like to avoid the trivial weak-coupling case
mentioned above, so in the following results we report here we take $g=1/\sqrt{2}$.  For this case, a $T$ of 200 realizes the interesting intermediate-data-length case.

\begin{figure}[H]
\center
\includegraphics[width=0.8\textwidth]{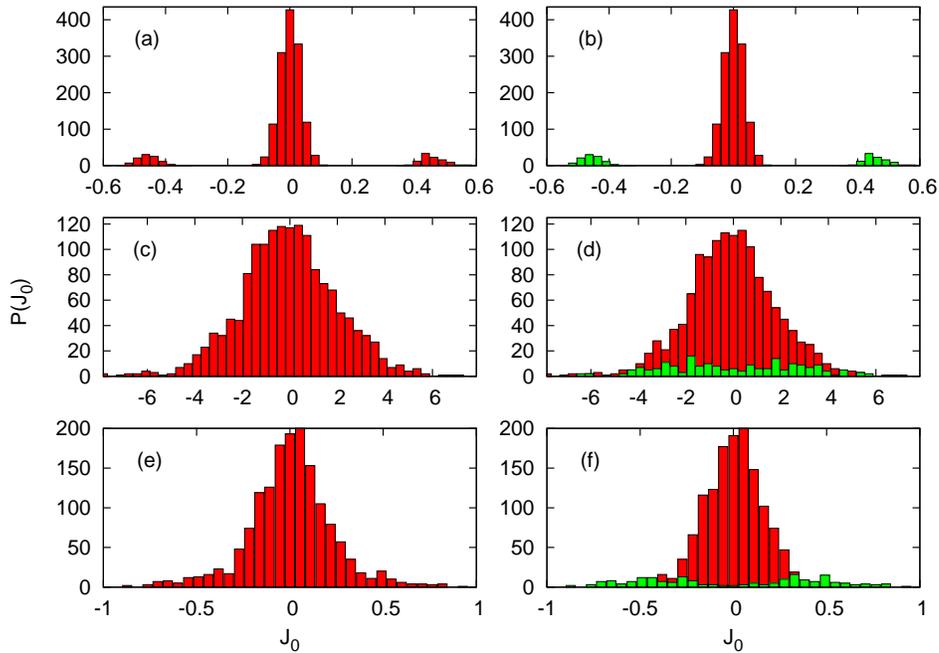}
\caption{Distribution of the inferred couplings without $L_1$ regularization, $g=1/\sqrt{2}$ for various data lengths. Top: $T = 2000$ updates/spin.  Middle: $T = 50$. Bottom: $T = 200$.
In each row, the left panel shows a histogram of the $J_{ij}$ obtained, and the right
panel shows these sorted according to whether the bond was present (green) or absent (red)
in the network that generated the data.}\label{Fig1}
\end{figure}

\begin{figure}
\center
\includegraphics[width=0.9\textwidth]{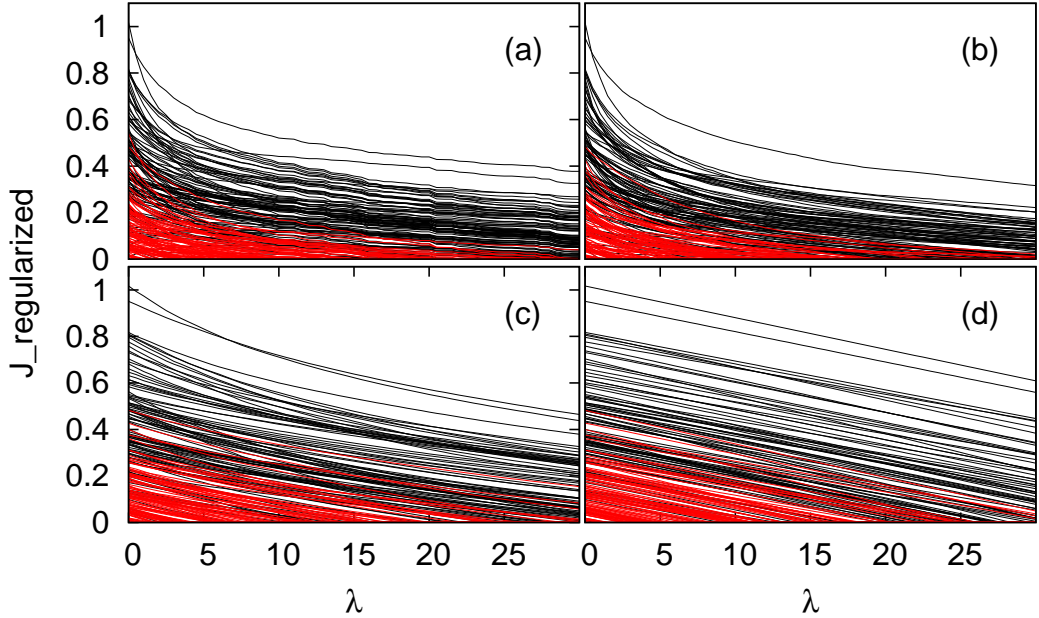}
\caption{Inferred couplings as functions of regularization coefficient $\lambda$ for four
methods: (a) full $L_1$ regularization using (\ref{eq:learning_rule}), (b)
integration of (\ref{eq:dJdlambda}), (c) integration of (\ref{eq:dJdlambda}) with diagonal
approximation of the inverse Fisher matrix, (d) linear extrapolation in $\lambda$ of the
curves in (c). Black lines represent bonds actually presents, while red lines represent
ones equal to zero in the network used to generate the data. We show equal number of red and black ones.}\label{Fig2}
\end{figure}

Based on $J$s inferred with $\Lambda_0=0$ as shown in figure \ref{Fig1}e and f, four pruning methods were employed. Figure \ref{Fig2} shows how the $J$s inferred by each method vary as the regularization coefficient $\lambda$ is increased. Here, we only show positive $J_0$s; graphs of the negative ones would look like the ones shown, reflected through the horizontal axis.  Bonds actually present in the model (a realization of
(\ref{eq:structure})) are plotted in black and bonds which are absent in red.

Figure~\ref{Fig2}a shows the $J$s inferred using exact learning with $L_1$ regularization
(\ref{eq:learning_rule}).  It is apparent that the pruning process for the case
shown here is not trivial in the way it would be in the weak-coupling limit: Some true
(black) bonds, for which rather small values were inferred at $\lambda=0$ because of
insufficient data, are ``rescued" (they fall off more slowly with $\lambda$ than red ones
with nearly the same initial inferred $J$s), and some spurious (red) bonds with high
inferred values at $\lambda=0$ are driven to zero faster than black ones with the same
initial inferred $J$s.  Thus, the red and black lines tend to be separated, and one can do
the pruning almost correctly just by turning $\lambda$ up until the desired number of
bonds have been removed.

Figure~\ref{Fig2}b shows the inferred $J$s using the quadratic expansion (\ref{eq:cost1}) in
the fashion described at the end of Sec.~\ref{sec4}. We call this ``approximation 1".
The qualitative features of figure~\ref{Fig2}a are apparently reproduced in this
approximation.

Figure~\ref{Fig2}c shows the result when off-diagonal elements of
$\left[C^{(i)}(\lambda)\right]^{-1}_{jk}$ are ignored in (\ref{eq:dJdlambda}). We
refer to this procedure as ``approximation 2".  The separation of red and black curves is
not as good in this case.  We also tried making a diagonal approximation of the Fisher
matrix itself, rather than its inverse: $C^{(i)}_{jk}$ by $C^{(i)}_{jj}\delta_{jk}$.
However, this gave much worse results (not shown) than making the diagonal approximation
on the inverse Fisher matrix.

In figure~\ref{Fig2}c, it is evident that the slopes of the $J_{ij}(\lambda)$ curves vary
rather slowly with $\lambda$.  Therefore, we also tried a linear extrapolation based on
the slopes of the curves in figure~\ref{Fig2}c at $\lambda = 0$. We denote this method as
``Approximation 3".  To the extent that this simple procedure works, one can identify the
nonzero bonds with very little computation:  One needs only to do the learning at
$\lambda = 0$ (to get the $J_{ij}(\lambda)$) and calculate the Fisher matrices (to get the
$dJ_{ij}/d\lambda$).  Figure~\ref{Fig2}d shows the result of this minimal algorithm.

For Approximation 3, the inferred $J$s that have been shrunk to zeros have no chance to be rescued again.  But for the other three approaches, the inferred $J$s for the positive ones (as shown in black lines) have that chance to be back again with increasing of $\lambda$. However, in the results presented in figure \ref{Fig2}, we haven't observe such phenomena.

One could also try similar linear extrapolation based on the initial slopes of the upper
panels of figure~\ref{Fig2}.  However, these curves show significant curvature for $\lambda
< 30$ or so, so the initial slopes are not good guides to the ultimate fate of the bonds
at large $\lambda$, and we do not present any results for these methods.

In what follows, we quantify the performances of these four pruning algorithms.  For the
three classes of bonds in the actual network, $-$, $+$ and 0, we can compute the empirical
classification errors. These errors can be either false positives (FP) (identifying a bond
which is really absent as present), or false negatives (FN) (identifying a bond which is
actually present as absent). In addition, a $+$ bond could be misclassified as $-$ or vice
versa, but this does not happen for the data length we are studying here.

\begin{figure}[H]
\center
\includegraphics[width=0.6\textwidth]{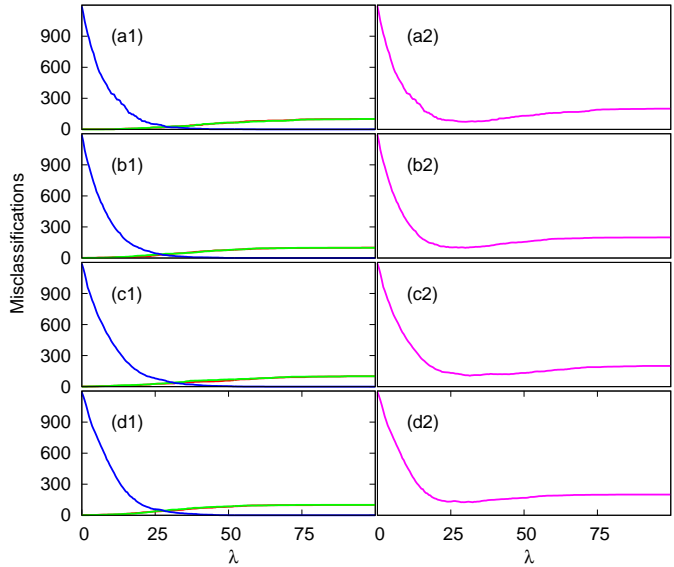}
\caption{Dependence of classification errors on $\lambda$. Left column: Number of
misclassified $-$, $+$, and $0$ (absent) bonds.  Numbers of false negatives for $-$s are
shown in green, for $+$s in red, and false positives for zero-bonds in blue.  Right
column: the sum of false negatives and false positives versus $\lambda$. Because the Js
are symmetrically distributed, red and green curves almost coincide, with mostly only the
green ones visible here. From top to bottom: full L$_1$ regularization and Approximations
1, 2, 3, respectively.}
\label{Fig3}
\end{figure}

At $\lambda=0$, where in general all bonds will be estimated to have nonzero values, there
will be no FNs and $N(N-c)$ FPs.   In the other limit $\lambda \to \infty$, all bonds will
be removed, so there will be $cN$ FNs and no FPs.   The empirical numbers of FPs and FNs
versus $\lambda$ are plotted in the left panels of figure \ref{Fig3}.   The total
misclassification error, i.e., the sum of the FPs and FNs (shown in the right panel of
figure \ref{Fig3}) has a minimum at $\lambda \approx 33.5$ for full L$_1$ regularization.
For Approximation 1 we find a minimum at $\lambda \approx 31.5$, while for Approximation 2
the minimum is at $\lambda \approx 30$, and for Approximation 3 it is at $\lambda \approx
24$.

\begin{figure}[H]
\center
\includegraphics[width=0.6\textwidth]{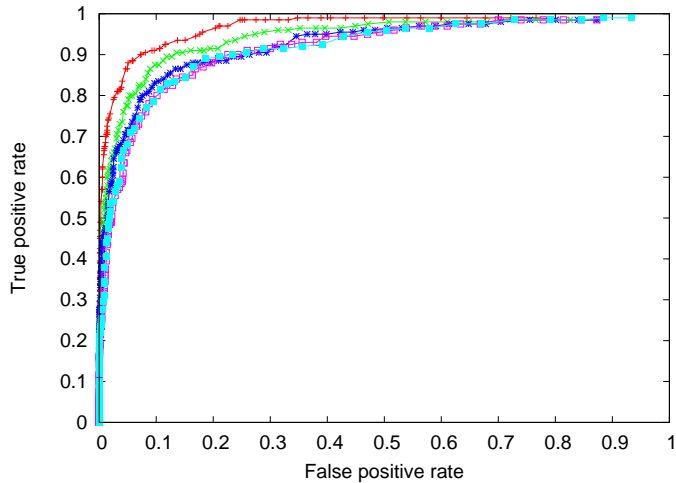}
\caption{ROC curves for full L$_1$ regularization, Approximations 1, 2, 3, and the J0-cut
method are shown in red, green, blue, pink and light blue, respectively.}\label{Fig4}
\end{figure}

In applications, FNs and FPs may not have the same cost associated with them: it may be
appropriate to weight the blue and green curves in the left-hand panels of figure~\ref{Fig3}
differently.  To compare algorithms in a more general way that is not specific to a
particular relative weighting, we calculate Receiver Operating Characteristic (ROC) curves
for them.   For a given $\lambda$, the false positive rate (FPR) is defined as the number
of FPs divided by the actual number of zero bonds, and the false negative rate (FNR) is
defined as the number of FNs divided by the number of actual number of non-zero bonds. A
true positive (TP) is the identification of a bond which is actually present as present,
and the true positive rate (TPR) is the number of TPs normalized by the actual number of
bonds present. It is equal to $1-\rm{FNR}$. The ROC curve is a plot of TPR versus FPR.
Each value of  $\lambda$ gives one point on the curve. In figure~\ref{Fig4}, we plot the ROC curves for all of our methods. We also measure the performance of the different methods quantitatively by defining an error measure, $\epsilon$:
\begin{eqnarray}\label{epsilon}
  \epsilon = 1-\rm{area~under~ROC~curve}.
\end{eqnarray}
The values of $\epsilon$s for full L$_1$ and Approximations 1, 2 and 3 are 0.03, 0.06, 0.08, 0.09 respectively. Thus, full L$_1$ algorithm performs best, followed by approximation 1. Approximation 2 works worse than them and it is only little better than approximation 3 for most values of $\lambda$, as can be seen in figure~\ref{Fig4}.

To establish a baseline for the goodness of our methods, we also performed a simple
pruning procedure that does not require any L$_1$ regularization calculation. For a given
cut value $\hat{J}$, we identify the bonds whose $J$s lie in the range
$[-\hat{J},\hat{J}]$ as absent and those outside that interval as present. The green $J$s
in figure~\ref{Fig1}f which lie within the interval are FNs and the red ones outside the
interval are FPs. Varying $\hat{J}$, we obtain an ROC curve. We refer to this procedure as
``J0-cut". The curve with light blue squares in figure~\ref{Fig4} is for it. The curve
nearly coincides with that for Approximation 3. Its $\epsilon$ is 0.09, the
same as that of Approximation 3. Thus, this trivial method works as well as Approximation 3.

\section{Effects of coupling strength $g$ on L$_1$ regularization}\label{sec6}
The above results were all obtained for $g=1/\sqrt{2}$. We are also interested in how the different regularization methods behave for other $g$ values. As we mentioned in section \ref{sec4} that in the weak couplings limit $g\rightarrow0$, the inverse of the Fisher information matrices for different $i$s are the same, equal to identity, thus the regularizations by Approximation methods will be equal to that of the trivial J0-cut.

\begin{figure}[H]
\center
\includegraphics[width=0.99\textwidth]{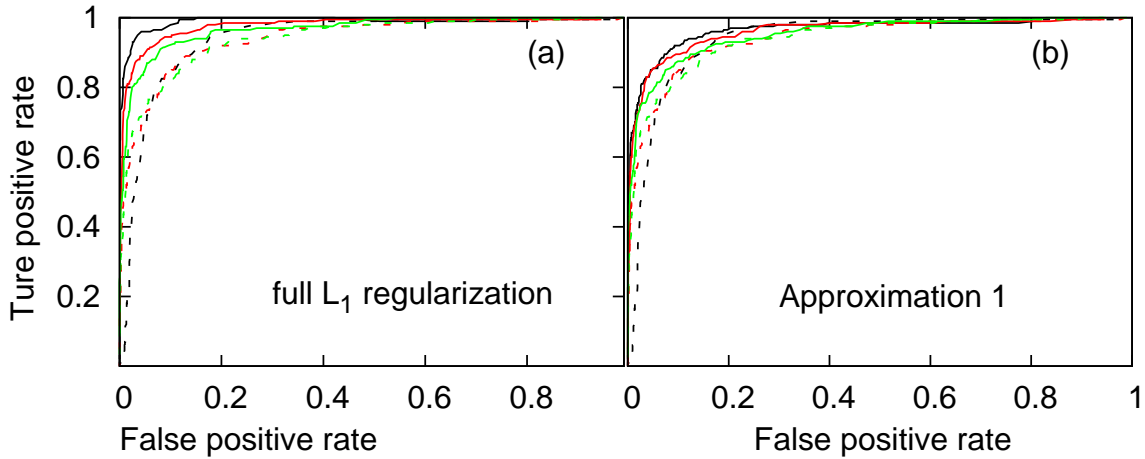}
\caption{ROC curves for full L$_1$ regularization (left, solid lines) and Approximations 1 (right, solid lines) with $g=\frac{1}{2}$, $\frac{1}{\sqrt{2}}$, $1$ respectively. The green lines for $g=\frac{1}{2}$, red for $g=\frac{1}{\sqrt{2}}$ and black for $g=1$. Corresponding dashed lines are for J0-cut method of these $g$s.}\label{Fig5}
\end{figure}

We repeat the calculation ROC curve for two other $g$s: 1 and 1/2. To get problem of the same level of difficulty, we first calculate the ROC curve by J0-cut method and make sure that the area under the curves are the same for each $g$.   A set of data lengths $L=NT$, for which the same areas under the curves can be obtained are found to be 11608, 8862 and 6730 for $g=1/2$, $1/\sqrt{2}$ and 1 respectively.   No bigger $g$ values are tested because they need shorter data length to get the same area, however, short data length increases the difficulties of the learning rule.   The ROC curves for all three $g$s are shown by the dashed lines in both figure \ref{Fig5}(a) and (b). The $\epsilon$ value are all around 0.94 for them.

With this stating point, we calculate the ROC curves for full L$_1$ regularization and Approximation 1 for all three $g$s.    As noted in figure \ref{Fig4}, the regularization by full L$_1$ and Approximation 1 have obviously better performances compared with that of the trivial J0-cut method, thus we next focus on this two methods to test whether regularization helps more at larger $g$. In figure \ref{Fig5}(a), the solid lines represent the ROC curves by full L$_1$ regularization.   The $\epsilon$ are 0.033 for $g=1/2$, 0.023 for $g=1/\sqrt{2}$ and 0.012 for $g=1$. Similarly, in figure \ref{Fig5}(b), the solid lines are for ROC curves by Approximation 1, with area 0.047, 0.039 and 0.035 for $g=1/2$, $1/\sqrt{2}$ and 1 respectively. As shown by the solid lines in both figure \ref{Fig5}(a) and (b), we can see that with increasing of $g$, the regularization methods work better. Both of them perform better than the trivial method, which is accordant with the results shown in figure \ref{Fig4}.

\section{Discussion}
\label{sec7}

We have studied the reconstruction of sparse asynchronously updated kinetic Ising
networks. With finite data length, simple maximization of the log likelihood of the system history will infer nonzero values to many bonds that are actually not present.  For large data length, this is generally not a problem, since the inferred bond distribution will consist of well-separated peaks. The ones with the smallest absolute values can then safely be identified as spurious and removed ``by hand''.  However, for smaller data lengths, these peaks can overlap strongly, and nontrivial methods are required to make an optimal pruning of the inferred coupling set.   Here we used L$_1$ regularization to do this, minimizing a cost function that includes the L$_1$-norm of the parameter vector as a penalty term.  We performed this minimization in four ways, one exact and the other three involving various degrees of approximation.

Calculations on a model network at intermediate coupling strength revealed that the exact L$_1$ regularization classified the bonds significantly better than a naive method based on retaining the strongest bonds.  Our Approximation 1 was somewhat worse than the exact algorithm, but still significantly better than the naive method.  Our other two approximations, obtained by successive simplifications of Approximation 1, however, did not perform measurably better than the naive way, as measured by the areas under their ROC curves.   These conclusions are general to various coupling strength we used.   The regularizations helps more with stronger coupling strengths.

This work is the first that we know of that takes a detailed look at how
L$_1$regularization works in the non-equilibrium model, by studying how bonds
are removed successively as the regularization parameter $\Lambda$ is increased.
Some insight into how this happens was made possible by studying the quadratic expansion of the cost function about its minimum, which also led to our relatively successful Approximation 1.   The process would have been more transparent if we could have made further simplifying approximations, as we did for Approximation 2, where we neglected off-diagonal elements of the inverse Fisher matrices. The fact that this approximation performed rather poorly (while Approximation 1 did quite well) indicates that the off-diagonal terms in (\ref{eq:dJdlambda}) are necessary, and we lack generic insight about them.

We performed our analysis here on a rather simple model network.  However, we expect that
our methods will be useful in analyzing date from a wide variety of biological, financial,
and other complex systems with sparse structure.

%This work can be considered as a pioneer that take a detailed looking at the shrinking ability of L$_1$ regularization. Based on comparison with other approximate regularization method, we found L$_1$ is quite intelligent in dealing with the present and absent bonds in regularizing process. This work is able to extend to deal with the real experimental data produced in biological or financial systems.

\section*{Acknowledgement} We are grateful to E. Aurell and M. Alava for useful
discussions about the work and Nordita and Niels Bohr Institute for hospitality. The work
of H.-L. Z. was supported by the Academy of Finland as part of its Finland Distinguished
Professor program, project 129024/Aurell.

\end{document}